\documentclass[journal=nalefd,manuscript=letter]{achemso}

\usepackage{amsmath, amsthm, amssymb}
\usepackage{graphicx}
\usepackage{xcolor}
\usepackage{xargs}
\usepackage[utf8]{inputenc}
\usepackage{ragged2e}

\usepackage{cleveref}
\crefname{equation}{Eq.}{Eqs.}
\usepackage{natbib}
\usepackage{subfigure}
\usepackage{graphicx}
 
\author{Guan-Wei Peng}
\affiliation{Department of Physics, National Taiwan University, Taipei 10617, Taiwan}
\alsoaffiliation{Research Center for Applied Sciences, Academia Sinica, Taipei 11529, Taiwan}
\altaffiliation{These authors contributed equally to this work.}

\author{Hung-Chin Wang}
\affiliation{Department of Physics, National Cheng Kung University, Tainan 70101, Taiwan}
\alsoaffiliation{Center for Quantum Frontiers of Research and Technology (QFort), National Cheng Kung University, Tainan 70101, Taiwan}
\altaffiliation{These authors contributed equally to this work.}

\author{Yu-Jie Zhong}
\affiliation{Department of Physics, National Cheng Kung University, Tainan 70101, Taiwan}
\alsoaffiliation{Center for Quantum Frontiers of Research and Technology (QFort), National Cheng Kung University, Tainan 70101, Taiwan}

\author{Chao-Cheng Kaun}
\affiliation{Research Center for Applied Sciences, Academia Sinica, Taipei 11529, Taiwan}
\email{kauncc@gate.sinica.edu.tw}

\author{Ching-Hao Chang}
\affiliation{Department of Physics, National Cheng Kung University, Tainan 70101, Taiwan}
\alsoaffiliation{Center for Quantum Frontiers of Research and Technology (QFort), National Cheng Kung University, Tainan 70101, Taiwan}
\alsoaffiliation{Academy of Innovative Semiconductor and Sustainable Manufacturing, National Cheng Kung University, Tainan 70101, Taiwan}
\email{cutygo@phys.ncku.edu.tw}

\title{Driving noncollinear interlayer exchange coupling intrinsically in magnetic trilayers}

\begin{document}

\begin{abstract}

Ferromagnetic side layers sandwiching a nonmagnetic spacer as a metallic trilayer has become a pivotal platform for achieving spintronic devices. Recent experiments demonstrate that manipulating the width or the nature of conducting spacer induces noncollinear magnetic alignment between the side layers. Our theoretical analysis reveals that altering the width of spacer  significantly affects the interlayer exchange coupling (IEC), resulting in noncollinear alignment. Through analytic and first-principles methods, our study on the Fe/Ag/Fe trilayer shows that at a specific width of the Ag spacer, the magnetic moments of side layers tend to be perpendicular. This alignment is mediated by Ag quantum well states, exhibiting spin spirals across the trilayer. Our results reveal that the noncollinear IEC offers a degree of freedom to control magnetic devices and boot spintronic technology with improved transport capabilities.

\begin{figure}[htb]
    \includegraphics[width=0.7\textwidth]{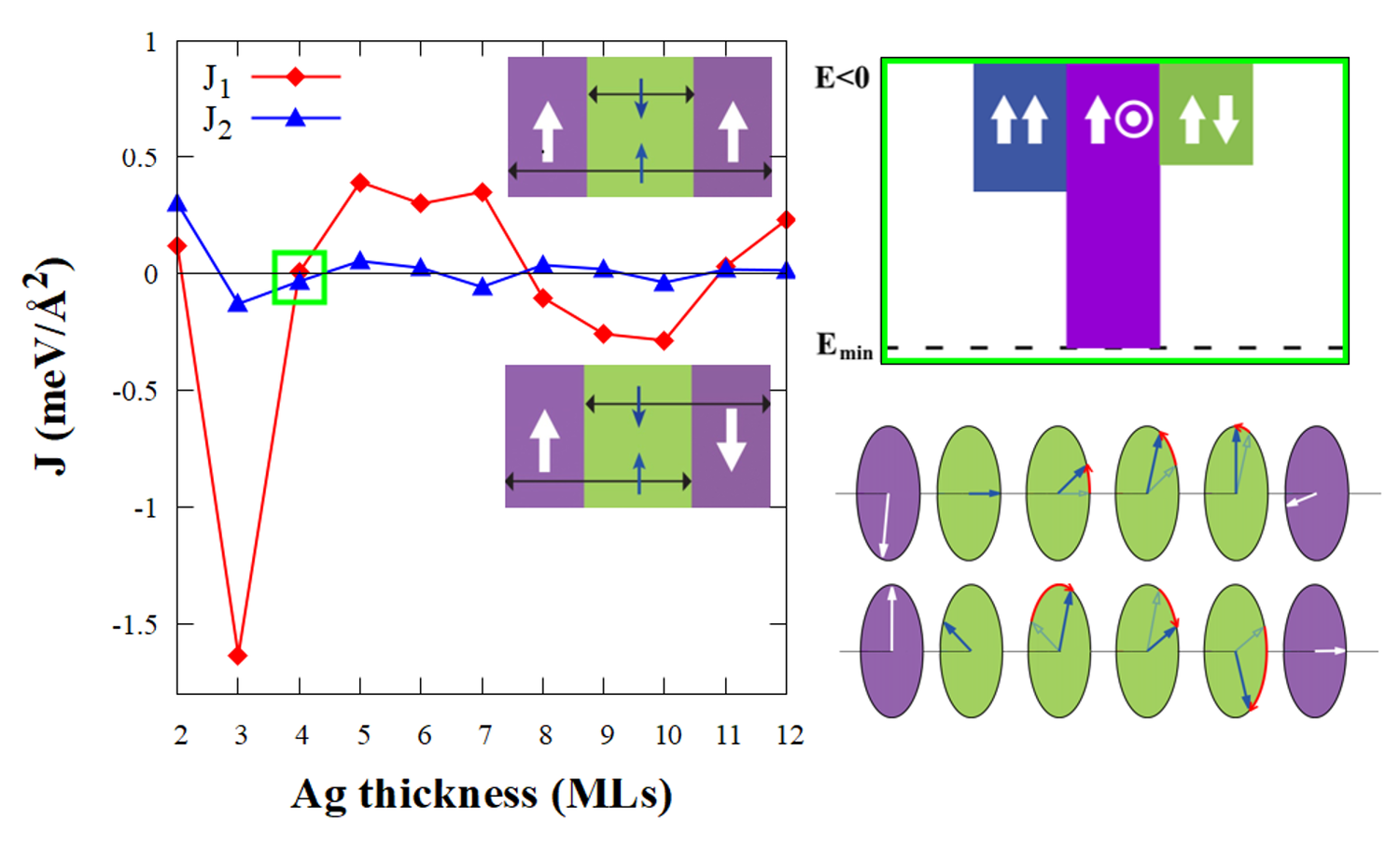}
    \centering
    \label{Abstractfig}
    \end{figure}

\end{abstract}
\maketitle

\section{Introduction}

A giant magnetoresistance (GMR) system, formed by ferromagnetic layers sandwiching a metallic spacer, has been widely used to fabricate various types of spintronic devices, such as magnetic random access memories, \cite{sousa2005nonvolatile,andrewd.kent2015new,bhatti2017spintronics} racetrack memories, \cite{parkin2008magnetic,parkin2015memory,yang2015domainwall} magnetic sensors, \cite{rolandweiss2013advanced,khan2021magnetic} and spin-torque oscillators \cite{kim2012spintorquea,chen2016spintorque}. Most of these devices consist of two or more magnetic layers separated by spacer layers that are used to mediate the IEC. \cite{kangwang2023spin,yang2017novel} Engineering the IEC to switch the ferromagnetic layers in the multilayer system between antiferromagnetic and ferromagnetic collinear alignments is a key in developing the memory and spintronic devices \cite{z.q.qiu2002quantum,rarembertduine2018synthetic,chcNPJ2017,chcPRB2017}.

In recent experiments of the Co/Ir/Co and Co/Ru/Co structures, however, it is found that not only the collinear, but also the noncollinear magnetic alignment between two Co layers can be created by either tuning the spacer width or alloying the Fe ions into the spacer. \cite{besler2023noncollinear,nunn2020control} This indicates that the components of the bilinear magnetic coupling ($J_1$) and the biquadratic magnetic coupling ($J_2$) coexist in the IEC and $J_2$ can be dominated by artificially altering the nature of GMR system. \cite{besler2023noncollinear}

Although the noncollinear alignment has been observed for more than 30 years in the GMR system including Co/Cu/Co \cite{heinrich1991magnetic} and Fe/Cu/Fe, \cite{ruehrig1991domain,a.azevedo1996biquadratic} it was confirmed that the interfacial defect and the quality of the film induce the extrinsic noncollinear IEC leading to such alignment. \cite{j.c.slonczewski1993origin,demokritov1998biquadratic,chc_roughness} In this work, we show that the intrinsic noncollinear magnetic alignment can appear in an ultrathin magnetic trilayer, consistent with recent experimental observations \cite{besler2023noncollinear,nunn2020control}.

We theoretically establish  that the noncollinear magnetic alignment in Fe/Ag/Fe trilayer stems from a competition between the bilinear coupling $J_1$ and the biquadratic coupling $J_2$ of IEC by using first-principles calculations and analytical techniques. By altering the width of Ag spacer to tune the IEC to turn off $J_1$, the perpendicular magnetic alignment appears in the Fe$_2$Ag$_4$Fe$_2$ trilayer. We further study the quantum well state (QWS) of noncollinear system and find that the spin of the Ag QWS must rotate either clockwise or counterclockwise along the direction perpendicular to the Fe/Ag interface. This indicates that the noncollinear magnetic structure can spatially rotate the carrier spin across the system, having rich application potential in developing spintronics.

\section{Methods}

\subsection{Quantum-well model}

\begin{figure}[tb]
\includegraphics[width=0.7\textwidth]{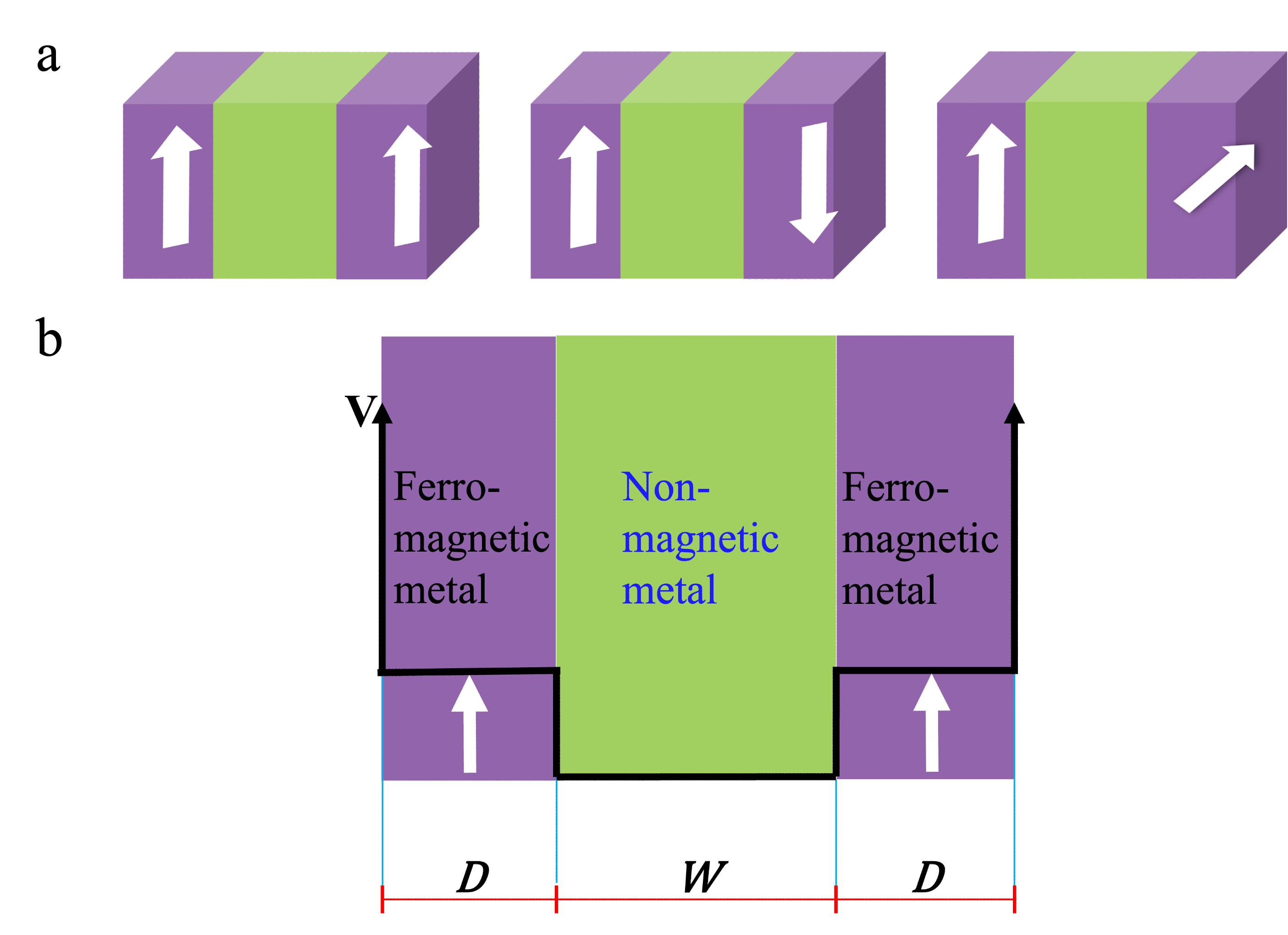}
\centering
\caption{
    \textbf{a, Schematic magnetic alignment between ferromagnetic layers.} The parallel , anti-parallel, and perpendicular alignments between the lateral ferromagnetic layers.
    \textbf{b, Schematic of the trilayer quantum well.}  
}
\label{fig1}
\end{figure}

To investigate a GMR system in the quantum-well model, we consider the metallic trilayered heterostructure 
consisting of the non-magnetic metallic Ag spacer with a width $W$ and   
Fe side layers with a width $D$, 
as shown in the Fig. 1a. 
Since the magnetic coupling stems from the energy variation in different magnetic configurations, we evaluate the total energy of the Fe/Ag/Fe trilayer in three systems referring to parallel, antiparallel, and  perpendicular alignments of the Fe side layers, respectively. 
The total energy of a GMR system is calculated by \cite{Switching}
\begin{equation}
E = \frac{U}{A} = \frac{E_F{k_F}^2}{4\pi ^2}\sum_{\sigma =\uparrow ,\downarrow }\sum_{n = 1}^{\infty} \int {d^2}k_{\|}\frac{\frac{E(\sigma ,n)}{E_F}+\frac{k_{\|} ^2}{k_F^2}-1}{\exp [(\frac{E(\sigma ,n)}{E_F}+\frac{k_{\|} ^2}{k_F^2}-1) \frac{E_F}{{k_B}T}]+1 },
\end{equation} 
where $E$ represents the total energy per unit area, $U$ is the total system energy, and $A$ is the system's cross-sectional area. 
For the carrier with spin $\sigma = \uparrow$ or $\downarrow$, the energy of its $n$-th band is $E(\sigma ,n)$, which depends on the magnetic structure of the system. 
The magnetic moments in the Fe layers leads to the Zeeman splitting of potential barriers for the majority and minority carriers \cite{Switching}.

Using a phenomenological approach \cite{demokritov1998biquadratic}, the total energy for the system per unit area is expressed as
\begin{equation}
E = -J_1(\vec{m}_1\cdot \vec{m}_2 )-J_2(\vec{m}_1\cdot \vec{m}_2 )^2, 
\end{equation}
where $J_1$ and $J_2$ are the coupling constant,
and $\vec{m}_i$ is the vector of normalized magnetic moment ($\left\lvert \vec{m}_i\right\rvert =1$). 
When the second-order interaction dominates ($2|J_2| > |J_1|$ ), the non-collinear system appears \cite{demokritov1998biquadratic,IECCH4}.

By using Eq. (1) to solve the system energy in different magnetic configurations,  we can estimate both bilinear and biquadratic components of IEC as 
\begin{eqnarray}
J_1 &=& E_{AP}-E_P \notag\\
J_2 &=& -\frac{E_{P}+E_{AP}}{2}+E_N,
\end{eqnarray}
(see Appendices Section A).
Here $E_{AP}$, $E_P$, and $E_N$ represent the energies per unit area of the antiparallel, parallel, and perpendicular configurations, respectively.



\subsection{First-principles calculations}
The Fe$_2$/Ag$_W$/Fe$_2$ sandwich structure was constructed with a bcc-like unit cell that consists of two ferromagnetic Fe monolayers (MLs) on each side and one metal Ag layer in the middle. 
The Ag layer thickens from 2 MLs to 12 MLs (see Appendices Section B). 
The vacuum for separating the Fe$_2$/Ag$_W$/Fe$_2$ is at least 18Å. The in-plane lattice constant was set to 2.89 Å, optimized by the Ag bulk.
The DFT calculations were performed by the Vienna Ab initio Simulation
Package (VASP) \cite{kresse1994normconserving,kresse1996efficiency,kresse1999ultrasoft}, with projector augmented wave (PAW) pseudopotentials \cite{kresse1999ultrasoft} and Perdew-Burke-Ernzerhof (PBE) generalized gradient approximation (GGA). 
The energy convergence criterion was set to less than $10^{-3}$ meV.
A $76 \times 76 \times 1$ k-point mesh centered at Gamma was used to sample the Brillouin zone. The energy cutoff for the plane-wave basis was set to 350 eV. 
The structurally optimized unit cells were obtained by the collinear calculation with the force convergence criterion of $10^{-1}$ meV/Å.
The noncollinear calculations for the three configurations (parallel, anti-parallel, and perpendicular alignments, see Fig. 1a) were performed without spin-orbital coupling. 
The estimated error in IEC was less than $10^{-2}$ meV/Å$^{-2}$ (see Appendices Section C).
In addition, we have calculated the magnetic anisotropy energies of our Fe$_2$/Ag$_W$/Fe$_2$ trilayer system (see Appendices Section D) to confirm that it prefers in-plane magnetization. \cite{chcNPJ2017, cinal_2003}

\section{Results and discussion}

\begin{figure}[tb]
    \includegraphics[width=0.9\textwidth]{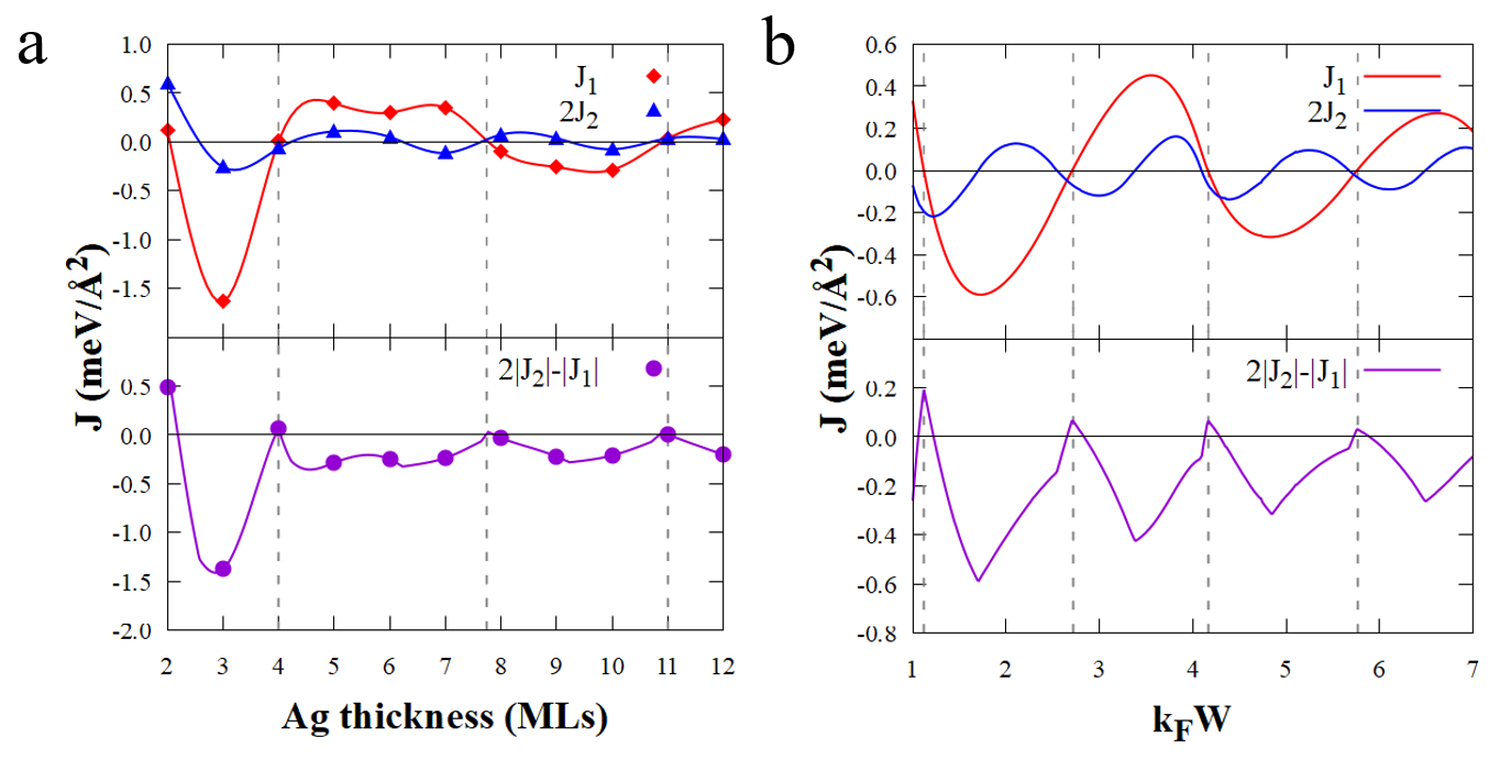}
    \centering
    \caption{
        \textbf{Interlayer exchange coupling according to the width of the spacer layer.} 
        \textbf{a}, The upper is the bilinear and biquadractic coupling strengths $J_1$ (red) and $J_2$ (blue), given by first-principles calculations. The line connecting the points is interpolated by the cubic spline. The lower shows the discriminant $2|J_2|-|J_1|$ (purple).
        \textbf{b}, Analytical results of $J_1$ (red), $J_2$ (blue) and $2|J_2|-|J_1|$ (purple).  The $k_F$ is the Fermi wavevector in the Ag layer and $W$ is the width of Ag spacer.
    }
    \label{fig2}
    \end{figure}

Using both first-principles and analytical calculations, the strengths of bilinear $J_1$ and biquadratic $J_2$ in the Fe/Ag/Fe trilayer with different spacer width are shown in Fig. 2.
In Fig. 2a, our noncollinear DFT calculations confirm that whereas the $J_2$ component dominates the magnetic coupling in the Fe$_2$Ag$_4$Fe$_2$ structure, the noncollinear magnetic alignment occurs between Fe layers.
In the upper of Fig. 2a, the diamonds (triangles) display the coupling strength $J_1$ ($J_2$) estimated by using Eq. (3) with the system energies obtained from first-principles calculations. 
Both $J_1$ and $J_2$  oscillate and decrease with the increasing of Ag spacer width.
The sign of $J_1$ directly indicates the magnetic alignment, in parallel or antiparallel orientations. 
Besides, the oscillation period of $J_1$ is almost twice than that of $J_2$, and the amplitude of $J_1$ is about 7 times greater than that of $J_2$. 

Based on Eq. (3), there are two requirements for the noncollinear magnetic alignment as the magnetic ground state in the GMR trilayer system: $2|J_2|-|J_1|>0$ and $J_2<0$.
In the DFT calculations shown in Fig. 2a, both Fe$_2$Ag$_4$Fe$_2$ and Fe$_2$Ag$_{11}$Fe$_2$ satisfy the requirements and prefer the noncollinear magnetic alignment.
It is worth nothing that three magnetic alignments in Fe$_2$Ag$_{11}$Fe$_2$ are almost degenerate as their energy difference is smaller than the numerical error.

To confirm that the noncollinear magnetic alignment is originated from the quantum resonance in the Fe/Ag/Fe trilayer, we estimate $J_1$ and $J_2$ by using analytical calculation based on solving the Schrödinger equation of Fe/Ag/Fe trialyer. The impact of Zeeman splitting in the system is considered as the spin-dependent potential barriers in Fe side layers. \cite{Switching} 
The analytical results shown in Fig. 2b consist with first-principles calculations in both amplitude and periods. 
Moreover, we confirm that the noncollinear coupling can occur at specific widths of spacer that dramatically reduce $J_1$, 
consistent with recent experiments in similar magnetic trilayer system. \cite{besler2023noncollinear,nunn2020control}

\begin{figure}[tb]
    \includegraphics[width=0.9\textwidth]{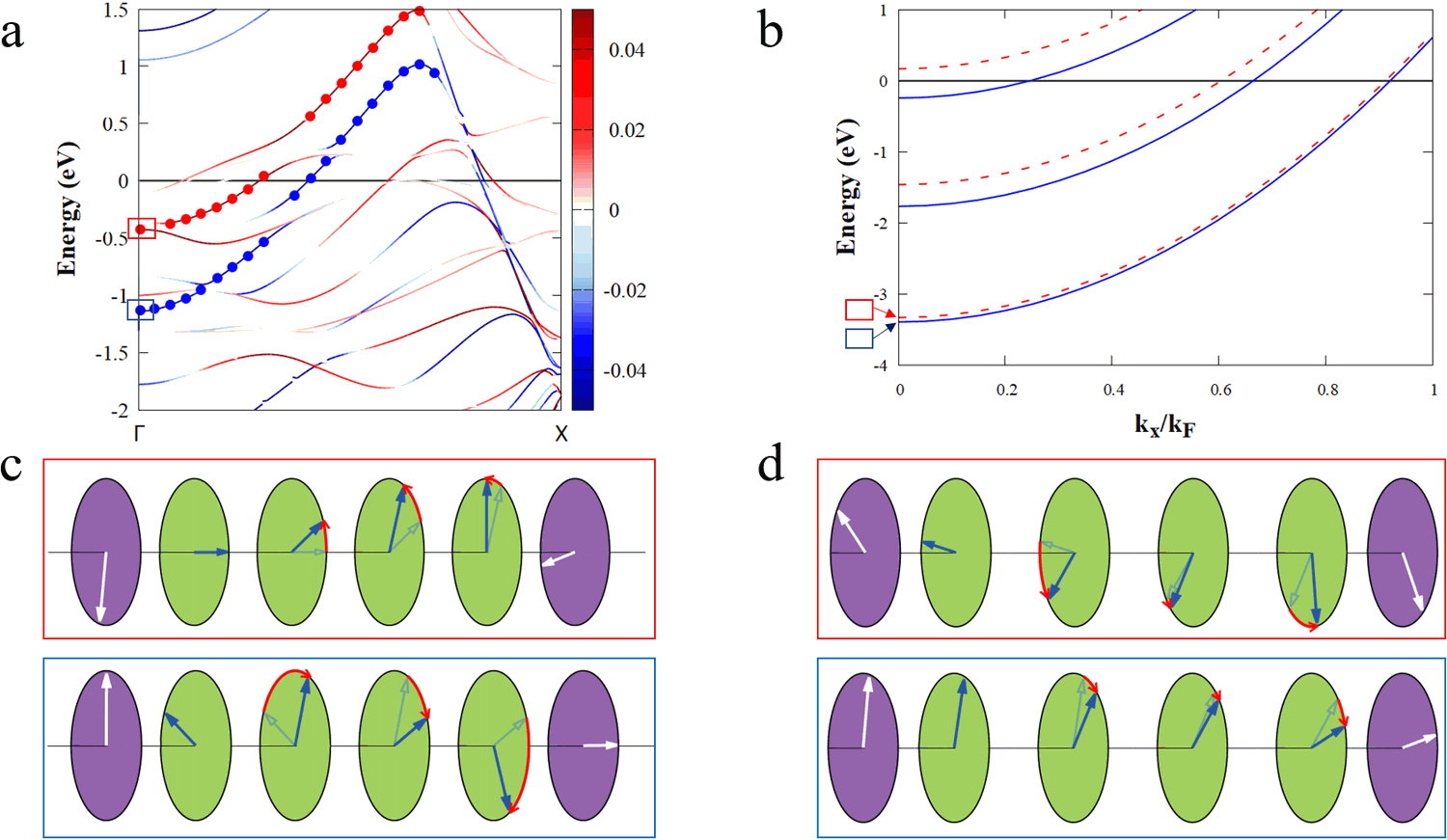}
    \centering
    \caption{
        \textbf{Band structure and spin rotation diagram of the system with perpendicular magnetic alignment.}
        \textbf{a}, The projected band structure of the noncollinear coupling Fe$_2$/Ag$_4$/Fe$_2$ by first-princles calculations. Only the orbitals in the transport direction ($z$)  ($p_z$ and $d_{z^2}$) of Ag atoms are projected. The color represents the x-component of the projected magnetization ($m_x$). The solid circles indicate the two bands of QWSs.
        \textbf{b}, The analytically calculated band structures. The blue (red) line denotes the QWS band with spin rotating clockwise (counterclockwise).
        The spin rotation diagrams from \textbf{c}, first-principles and \textbf{d}, analytical calculations, corresponding to the boxes in Figs. \textbf{a} and \textbf{b}. The arrow is the normalised spin with transverse direction ($x$ and $y$), and the horizontal direction is the transport direction. The colors of the solid circles represent different layers, green for the spacer layer (Ag) and purple for the magnetic layer (Fe).
        }
    \label{fig3}
    \end{figure}

To understand the nature of electronic states in a noncollinear GMR system, 
we calculate both band structures and spacial spin orientations of Ag QWSs in the Fe$_2$/Ag$_4$/Fe$_2$ trilayer with noncollinear magnetic alignment.
Figure 3a shows the band structure projected into the $p_z$ and $d_{z^2}$ orbitals of the Ag atoms, along the transport direction. 
The color represents the x-component of the projected magnetization ($m_x$). 
At the $\Gamma$ point of  two bands marked with circles are the electric-like QWSs. \cite{bruno1991oscillatory,kawakami1999determination,chang2015engineering} 
The projected magnetization of these bands at the $\Gamma$ point on each atom (referring to the spin of QWSs) rotates around the transport axis, shown in Fig. 3c.  
It is worth noting that these unique bands always appear in pairs and the configuration of band splitting is similar to the Zeeman splitting.
These two rotations, clockwise $270^\circ$ and counterclockwise $90^\circ$, are exactly the two ways of a $90^\circ$ rotation when less than a turn. 

We also estimate the bands and the spin of QWSs in a simple analytical model (Figs. 3b and 3d), and the results agree qualitatively with those from DFT calculations (Figs. 3a and 3c).
In Figs. 3d and 3c, we confirm that the spin orientation in Ag QWSs can rotate spatially in a GMR system with noncollinear magnetic alignments, even without the spin orbital coupling.
This indicates that the noncollinear IEC not only adds a degree of freedom in tuning the magnetic structure of GMR system, but also provides a way to rotate the spin of  carriers in real space.
The detail information of bands and QWSs obtained by our analytical calculations is provided in Appendices Section E.
Besides, we have confirmed that the spacial spin oscillation of Ag QWSs and $J_1$ have the same period and this feature can be kept in a thicker GMR system (see Appendices Section F).  
Our results thus explain the observed spacial Ag spin polarization in the similar GMR system \cite{referee_prl}. 

\begin{figure}[tb]
    \includegraphics[width=0.5\textwidth]{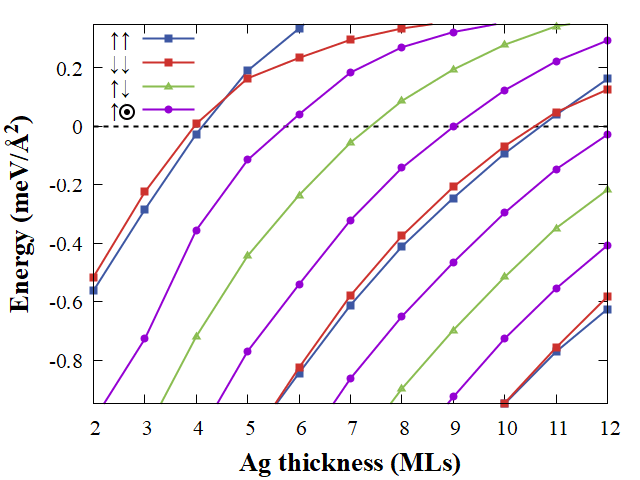}
    \centering
    \caption{
        \textbf{Energies of quantum well states versus the width of Ag spacer (D) in Fe$_2$/Ag$_W$/Fe$_2$.}
        The blue and red squares represent the majority and minority QWSs in the parallel alignment of the Fe layers. The green triangles correspond to the QWSs in the anti-parallel alignment of the Fe layers. The purple circles denote the QWSs in perpendicular alignment of the Fe layers. The lines connect the energy points within the same quantum number.
    }
    \label{fig4}
    \end{figure}

To correlate IEC with quantum well resonances in GMR systems, we plot the energy minimum of QWS bands in different magnetic configurations in Fig. 4.
When the energy minimum pass Fermi level (see flat dashed line in Fig. 4), the quantum resonance occurs in the GMR system.  
For the GMR system with parallel magnetic alignment, for instance, the quantum resonance near Fermi level occurs when the width of Ag layer is around 4 and 11 MLs (see blue and red squares  in Fig. 4). 
The period of quantum resonances is around 7 ML, directly respected to that of $J_1$ in IEC (Fig. 2a), 
consistent with the previous works of collinear IEC \cite{bruno1991oscillatory,kawakami1999determination,chang2015engineering}.

The IEC and quantum resonance share the same period in space, occurring in $J_2$ as well.  
Since the quantum resonance in a GMR system with the perpendicular alignment appears as the width of Ag layer is around 6 and 9 MLs (see purple circles in Fig. 4), its period is approximately 3 MLs, 
consistent with the period of $J_2$ (see up panel in Fig. 2a). 
Based on the direct correlation between Fig. 2 and Fig. 4, we  confirm that the perpendicular alignment in a GMR system can be intrinsically driven by controlling the width of GMR system.

\section{Conclusion} 

To sum up, we theoretically establish that the magnetic IEC can intrinsically drive perpendicular magnetic alignment between ferromagnetic side layers in an ultrathin GMR system.
The nonlinear IEC can be triggered by selecting a specific width of spacer layers where the linear term of IEC almost vanishes, consistent with recent experimental observations in GMR systems. \cite{besler2023noncollinear,nunn2020control}
Moreover, we find that the intrinsic IEC not only adds more degree of freedom to select magnetic structures, but also induces the spacial spin orientations of the itinerary carriers in GMR systems, as is observed in the similar system \cite{referee_prl}.
These findings can deepen the understanding of nonlinear coupling and have potentials for developing advanced spintronic devices.

\begin{acknowledgement}
This work was supported in part by the Higher Education Sprout Project, Ministry of Education to the Headquarters of University Advancement at the National Cheng Kung University (NCKU). We acknowledge the financial support by the National Science and Technology Council (Grant numbers NSTC-112-2112-M-006-026-/112-2112-M-001-072-/112-2112-M-006-015-MY2) and National Center for High-performance Computing for providing computational and storage resources. C.H.C. thank the support from the Yushan Young Scholar Program under the Ministry of Education in Taiwan.\end{acknowledgement}

\bibliography{GMR_multilayer_IEC.bib}


\end{document}